# Image Search Reranking

**V Rajakumar, M.Tech (IT)**
*Associate Professor, CSE Dept, ASTRA*

**Vipeen V Bopche, M.Tech**
*IT Dept, ASTRA, Bandlaguda*

**Abstract: -** *The existing methods for image search re ranking suffer from the unfaithfulness of the assumptions under which the text-based images search result. The resulting images contain more irrelevant images. Hence the re ranking concept arises to re rank the retrieved images based on the text around the image and data of data of image and visual feature of image. A number of methods are differentiated for this re-ranking. The high-ranked images are used as noisy data and a k-means algorithm for classification is learned to rectify the ranking further. We are study the affectability of the cross validation method to this training data. The preeminent originality of the overall method is in collecting text/metadata of image and visual features in order to achieve an automatic ranking of the images. Supervision is initiated to learn the model weights offline, previous to reranking process. While model learning needs manual labeling of the results for a some limited queries, the resulting model is query autonomous and therefore applicable to any other query .Examples are given for a selection of other classes like vehicles, animals and other classes.*

## 1. INTRODUCTION

Information retrieval is the process of obtaining information resources relevant to data need from a collection of information materials. Searches can be based on metadata of image or on full-text or content-based indexing.

An IRS begins when a user enters a query into the system. Queries are conventional statements of information requires, for example search strings in web search mechanisms. In information retrieval a query does not unusually identify a single object in the collection.

Image search engines apparently provide an off head route, but currently are limited by poor precision of the returned images and also restrictions on the total number of Images provided. With Google Image Search, the precision is as small as 32 percent on one of the classes tested here (shark) and averages 39 percent, and downloads are confined to 1,000 images.

The text based image contains relevant and irrelevant image results. All of the existing reranking algorithms require earlier assumption regarding the congruity of the images in the opening, text-based search result.

In proposed system objective of this work is to retrieve a more number of images for a specified object category from the browser. A multimodal approach employing text, metadata, and optical features is used to congregate many high-quality images from the internet. Candidate images are acquired by a text-based online search querying on the object identifier. The task is then to remove unwanted images and re-rank the remaining. The images are re-ranked formed on the text arounding the image and metadata features of images. A number of methods are differentiated for this re-ranking. The high-ranked images are used as training data and k-means clustering classifier is learned to set right the ranking further.

## 2. LITERATURE SURVEY

2.1 Supervised Reranking for Web Image Search

The existing image search engines, including Yahoo, Google, and Bing, recovers and rank images mostly based on the textual information associated with an image in the arranged web pages, such as the name of image and the a rounding text. While text-based image ranking is often effective to search for related images, the precision of the search results limited by the dissimilarity between the true relevance of an image and its relevance implicit from the associated textual descriptions.

*SVM algorithm*

The objective of the learning-to-rerank task is to estimate the parameters by minimizing a loss function. Methods that can be used for this function. Ranking SVM is a classic.

Algorithm applied in learning-to-rank, purpose differs in the design of the loss where the loss function is defined as a combination of the prediction loss and the regularization term. Support vector machines are supervised learning models with associated learning algorithms that analyze data and recognize patterns, used for classification and reverting analysis. The SVM takes a set of input information and projects, for each given input, which of two possible classes forms the result, making it a certainty binary linear classifier. Given a set of training, each marked as belonging to any one classes, a SVM training algorithm constructs a model that assigns new examples into one class or the other.

*PRF*

Relevance feedbacks a feature of some IR systems. The idea behind relevance feedback is to take the results that are initially returned from a given query and to use related data about whether or not those results are relevant to perform a new query. We can usefully differentiate between three types of feedback implicit feedback, explicit feedback, and pseudo feedback.

2.2 OPTIMOL: automatic Online Picture collection via Incremental Model Learning

OPTIMOL approaches the problem of learning object categories from online image searches by addressing model learning and searching jointly. OPTIMOL is a repetitive model that updates its model of the target object class while concurrently retrieving more related images. Important task in image understandings object recognition, in particular general object distribution. Analytical to this problem are the matter of learning and dataset. Bounteous data helps to train a robust identification system, while a good object classifier can succor to collect a more number of images.

2.3 Overview of the proposed prototype-based visual reranking

The proposed prototype-based reranking method inheres of an online and an offline step. In the online part, when a textual query is given in to the image search engine, initial search is performed using any in fashion text-based search technique. Then, visual prototypes are created and for each and every prototype a Meta reranker is build. The offline component is affectionate to learning the reranking model from human-labeled data like tags, comments. Since the learned model will be used the text-based search results for reranking, the training set is build from these results through the following way. After the training data is collected, we can calculate the score vector Meta rerankers, as per in the online part, for each and every image and the corresponding query. prototype-based reranking model, which builds meta rerankers corresponding to visual feature representing the textual query and learns the weights of a linear reranking model to combine the results of meta rerankers and create the reranking score of a given image taken from the earlier text-based search result.

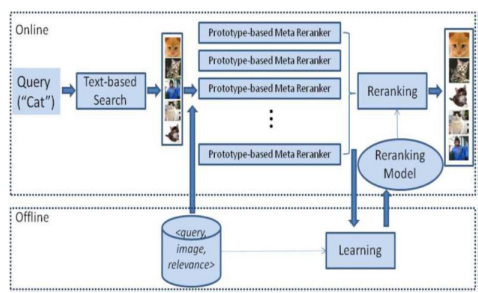

Fig 1: Overview of the prototype-based reranking

## 3. SYSTEM ANALYSIS

*Data Collection*

We analyze three different kinds of approaches to downloading images from the internet by IO streaming. The first approach, named Web Search (Google, yahoo, Bing), presents the query word to Google search and all images which are related or none related that are linked within the returned Web pages link are downloaded. Google limits the number of retrieved Web pages to 1,044, but some of the Web pages contains more than one or two images, so in this kind, 1000 of images are obtained. Google search limits the number of retrieved images to 1,000, but here, each of the obtain images is treated as a seed"—further images are downloaded from the Webpage by IO streaming where the seed image originated. The Google Images, contains only the images directly returned by Google search. The query can contains a single word or more specific descriptions such as "penguin animal" or "penguin OR penguins." Images are lesser than 120_120 are discarded. In addition to the images, texts around the image HTML tags are downloaded, with other metadata of image such as the image filename.

*Removing Drawings and Symbolic Images*

Since we are mostly interested in building databases for natural image identification, we ideally would like to remove all abstract images from the downloaded images. Separating abstract images from all others automatically is very challenging for classifiers based on visual features.

Instead, we outfit the easier visual task of removing drawings and symbolic images. These includes: plots, comics, maps, charts, graphs, drawings, and sketches, where the images can be fairly simply characterized by their visual features

*Results for Textual ether Visual Image Ranking*

In this section, we evaluate different combinations of training and testing. If not stated otherwise, the text+vision system was used. For each choice, five different arbitrary selections are made for the sets used in the 10-fold cross validation and mean and quality deviation is reported. The clear enhancement brought by the visual classifier over the text-based ranking for most classes is obvious.

There are two possible cases that can happen:

A parameter setting that over fits to the training data. This problem is observed on the validation set due to a low precision at 15 percent recall.

All images (training and validation sets) are classified as background. This leads to bad, but measurable, performance as well.

Here, we describe a slight adjustment to this method, which ignores "difficult" images. We then use the "good" parameter settings to train and classify all images. We are able to eliminate "intermediate" images, i.e., images that are not grouped as positive or negative images in the majority of cases. We guess that those images are difficult to classify.

## 4. SYSTEM DESIGN

MODULES OF THE SYSTEM

There are five different types of modules in this project that are as follows.

4.1 Query Image

When an image search in search engines, that same images are downloaded in that time, meanwhile among them there is an unrelated images are also obtained. However, producing such databases consists of a many number of images and with precision is still a backbreaking manual task.

Generally search engines apparently provide a simple route. For this type of acquiring images can be filter.

4.2 Download Associate Images

We compare three different kinds of approaches to downloading images from the Web by IO streaming.

The first approach, named Web Search, give the query words to Google search and each and every images that are linked within the returned Web pages are downloaded by IO streaming. Google limits the number of go back Web pages to 1,000 and more but many of the Web pages contains multiple images, so in this way, 1000 of images are obtained.

The next approach, Image Search, starts from Google image search or other web search. Google image search controls the number of returned images to 1,000, all the returned images is treated as a "seed"—further images are downloaded from the Webpage by using IO streaming where the seed image originated.

The last approach, Google Images includes only the images directly collect by Google image search or other search engine. The query can made up of a single word or specific descriptions such as "penguin animal" or "penguin OR penguins." Images lesser than 122 _ 122 are thrown. In addition to the images, texts around the image HTML tags are downloaded, together with other metadata of image such as the image filename.

4.3 Apply Re-ranking model

Now describe the re-ranking of the returned images based on metadata of image and text surrounding of image alone. We extend and follow the method proposed by using a set of textual features which presence is great indication of the image content.

The aim is to re-rank the searched images. Each feature is used as binary: "True" if it contains the query text and "False" otherwise. To re-rank images for one fix class, we do not employ the whole images for class.

4.4 Filtering Process

The re-ranker on average, performs well, and significantly increases the precision up to quite a high some level. To re-ranking the filtered images, we have applied the text vision system to all images downloaded from internet for one specific class, i.e., the drawing and symbolic images were included. Record that the performance is comparable to the case of filtered images. The well formed visual model is strong enough to delete the symbolic and drawings images during the ranking process. The filtering is necessary to train the visual classifier and is not required to rank new images.

## 5. SYSTEM ARCHITECTURE

The systems architect establishes the basic structure of the system, defining the important core design features and elements that provide the framework. The systems architect provides the architects view of the users' vision.

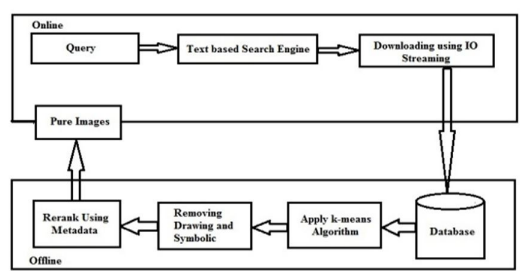

Fig 2: System Architecture

In above diagram proposed system consists of an online and offline step. In the online part, when a textual query is given to the image search engine, initial search is performed using any in fashion text-based search technique. Then download the retrieve images using IO streaming. In offline part, Taking top rank image as a dataset. And store images in database on local system. Applying k-means algorithm on the downloaded images and removing drawing and symbolic images after that reranking the remaining images using metadata of the images. Getting pure images on the web browser as output.

## 6. IMPLIMENTATION

Submit query to Google search engine, Here we are submitting query to the Google search engine for this we are connecting the Google search engine by the help of web browser and submitting the query.

Class URL, it represents a Uniform Resource Locator, a pointer to a "resource" on the WWW. A resource can be something as simple as a file or a directory, or it can be referred to a more complex object, such as a query to a database or to a search engine.

URLConnection, The URLConnection class contains many methods that let you communicate with the URL over the N/W. URLConnection is an HTTP-centric class; that is; many of its methods are useful.

Returns a URL Connection object that represents a connection to the remote object referred to by the URL. A new connection is created every time by calling the openConnection method of the protocol handler for this URL.

SetRequestProperty, Sets the general request property. If a property with the key exists before only, overwrite its value with the new value.

BufferedReader reads text from an input type character. Input stream, buffering characters so as to provide for the efficient reading of arrays, characters and lines.

File Creates a new File instance by converting the given pathname string into a pathname. If the given string is the empty string, then the output is the empty abstract pathname.

Creating Database Table ,Here we are creating a data base table for storing image url , height, width, size and start number to the image.

Imageurl, in this column image url is store.

Height, In this column we are storing height of the images, by using getheight() method which is in java package method. This method determines the height of the image if height is not yet known this method return -1, otherwise height of the image.

Width, in this column we are storing Width of the images, by using getWidth () method which is in java package method. This method determines the Width of the image if Width is not yet known this method return -1, otherwise Width of the image.

Sz, in this column we are storing size of image. Digital images are always square or rectangle. They are constructed of pixels. There is a vertical number and a horizontal number of pixels. When you multiply those two numbers, you get the pixel size of the image.

Height, by using getheight () method which is in java package. This method determines the height of the image if height is not yet known this method return -1, otherwise height of the image.

Width, by using getWidth () method which is in java package. This method determines the Width of the image if Width is not yet known this method return -1, otherwise Width of the image.

Sz, Digital images are rectangles. They are constructed of pixels. There is a vertical number and a horizontal no of pixels. When you multiply horizontal and vertical pixels numbers, you obtain the pixel size of the image.

Skew, the lines of text within images of scanned documents can usually be brought into nearly upright orientation with a rotation of a multiple of 90 degrees. If the required rotation is 0 or 180 degrees, the images are in portrait mode; if 90 or 270 degrees it is in landscape mode. After such rotation, the text lines (if they exist) usually have a small amount of skew.

Energy, we define an energy that would capture the solution we desire and perform gradient descent to computer to its minimum value, resulting in a solution for the image distribution. Finding height, width and size of images Finding energy of images, Finding entropy of images, finding skew of images extracting the pure images as output, here we are retrieving the pure images as output.

## 7. CONCLUSIONS

We proposed an online image search and reranking framework, in which build Meta reranker corresponding to visual prototype like size of the image, height of the image, width, energy, skew, entropy, and other visual feature representing the textual query and learns a linear reranking model to combine the output of individual meta reranker and generate the reranking score of the consisting image taken from the earlier text based search result. Some models are process query depended but our model is query independent way requiring only a limited labeling effort and being able to large range of images. Our model improves the performance over the text based search result by collecting prototypes and textual ranking features. A main extension of the approach describe in this paper would be to apply method to study concept models from image search engines in semi automatic fashion. Our model increasing precision and give the accuracy. Our results show that this assumption can be regarded valid in a general case, still diversions from this expectation can occur for individual queries. Hence, we could work on improving the proposed reranking model to make it more query-adaptive.

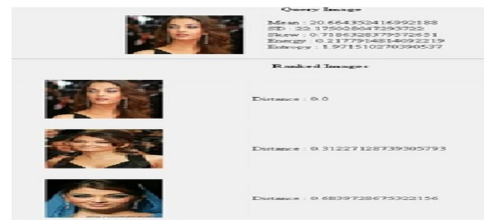

First Author: **V Raja Kumar**, Associate Professor, Aurora's Scientific Technological and Research Academy, completed his M. Tech (IT) from JNTU, Hyderabad. He published several research papers in the field of Information Technology and Network Security. He attended the conference of management skills program the head of the institution at Administrative Staff College of India. His areas of research are Information Technology, Network Security.

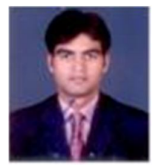

Second Author: **Vipeen V Bopche** received his Bachelor of Engineering (BE) degree from Sant Gadge Baba Amravati University in 2011. He is currently pursuing M. Tech. in

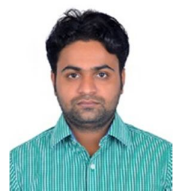

Information Technology, from ASTRA, Bandlaguda affiliated from Jawaharlal Nehru Technological University, Andhra Pradesh. His research interests are in Data Mining, Cloud Computing, and Image Processing.